\documentclass[11pt]{article}
\usepackage{setspace}
\usepackage{epsfig,subfigure}
\usepackage{amssymb}
\usepackage[fleqn]{amsmath}
\usepackage{arydshln}

\usepackage{color}
\newcommand{\define}{\stackrel{\triangle}{=}}
\def\QED{\mbox{\rule[0pt]{1.5ex}{1.5ex}}}

\usepackage{graphicx}
\usepackage{color}

\usepackage{amssymb}
\usepackage{amsmath,amsfonts,amssymb}
\usepackage{verbatim}
\usepackage{stfloats}
\usepackage[bookmarks=false]{}

\newtheorem{theorem}{Theorem}

\newtheorem{lemma}{Lemma}

\newcommand\blfootnote[1]{%
  \begingroup
  \renewcommand\thefootnote{}\footnote{#1}%
  \addtocounter{footnote}{-1}%
  \endgroup
}

\begin{document}
\date{}
\title{Multiround Private Information Retrieval: \\ Capacity and Storage Overhead
}
\author{ \normalsize Hua Sun and Syed A. Jafar \\
}

\maketitle

\blfootnote{Hua Sun (email: huas2@uci.edu) and Syed A. Jafar (email: syed@uci.edu) are with the Center of Pervasive Communications and Computing (CPCC) in the Department of Electrical Engineering and Computer Science (EECS) at the University of California Irvine. 
}

\begin{abstract}
The capacity has recently been characterized for the private information retrieval (PIR) problem as well as several of its variants. In every case it is assumed  that all the queries are generated by the user simultaneously. Here we consider multiround PIR, where the queries in each round are allowed to depend on the answers received in previous rounds. We show that the capacity of multiround PIR is the same as the capacity of single-round PIR (the result is generalized to also include $T$-privacy constraints). Combined with previous results, this shows that there is no capacity advantage from multiround over single-round schemes, non-linear  over linear schemes or from $\epsilon$-error over zero-error schemes. However, we show through an example that there is an advantage in terms of storage overhead. We provide an example of a  multiround, non-linear, $\epsilon$-error PIR scheme that requires a strictly smaller storage overhead than the best possible with single-round, linear, zero-error PIR schemes.  
\end{abstract}

\newpage
\allowdisplaybreaks
\section{Introduction}

Private information retrieval (PIR) \cite{PIRfirst, PIRfirstjournal} is one of the canonical problems in theoretical computer science and cryptography. The PIR setting involves $K$  messages that are assumed to be independent, $N$ distributed databases that are replicated (each database stores all $K$ messages) and non-colluding (the databases do not communicate with each other), and a user who desires one of the $K$ messages. A PIR scheme is any mechanism by which a user may retrieve his desired message from the databases privately, i.e., without revealing any information about which message is being retrieved, to any individual database. An information theoretic formulation of PIR guarantees the user's privacy even if the databases are computationally unbounded.\footnote{There is also a widely studied cryptographic formulation of PIR, where the user's privacy is guaranteed only against computationally bounded databases \cite{CPIR, William, Yekhanin}.}
The ``rate" of a PIR scheme is defined as the ratio of the number of bits of desired information to the total number of bits downloaded by the user from all the databases. The supremum of  achievable rates is defined to be the capacity of PIR. For $K$ messages and $N$ databases, the capacity of PIR was characterized recently in \cite{Sun_Jafar_PIR} as
\begin{eqnarray}
C&=&\left(1+1/N+1/N^2+\cdots+1/N^{K-1}\right)^{-1}\label{eq:C}
\end{eqnarray}

The capacity  has also been determined for various constrained forms of PIR such as LPIR \cite{Sun_Jafar_PIRL} -- where message \emph{lengths} can be arbitrary, TPIR \cite{Sun_Jafar_TPIR} -- where any set of up to $T$ databases may collude, RPIR \cite{Sun_Jafar_TPIR} -- where \emph{robustness} is required against unresponsive databases,  SPIR \cite{Sun_Jafar_SPIR} -- which extends the privacy constraint \emph{symmetrically} to protect both the user and the databases, MDS-PIR \cite{Banawan_Ulukus} and MDS-SPIR \cite{Wang_Skoglund} -- variants of PIR and SPIR, respectively, where  each message is separately MDS coded.\footnote{As a caveat, we note that separate MDS coding of each message is a restrictive assumption. Consider the setting with $K = 2$ messages, $N = 3$ databases and the storage size of each database is equal to the size of one message. If separate MDS codes are employed for each message, then the maximum rate (capacity) is equal to $3/5$ \cite{Banawan_Ulukus}. However, Example 2 in \cite{Chan_Ho_Yamamoto} shows that rate $2/3$ $(> 3/5)$ is achievable with a storage code that jointly encodes both messages.}

A common theme in these results is that there is no capacity advantage of non-linear schemes over linear schemes, or of $\epsilon$-error schemes over zero-error schemes. This is a matter of some curiosity because the necessity of non-linear coding schemes has often been a key obstacle in network coding capacity problems \cite{dougherty1, CG_Duality, Rouayheb_Sprintson_Georghiades, Blasiak_Kleinberg_Lubetzky_2011}, and the capacity benefit of $\epsilon$-error schemes over zero-error schemes for network coding problems in general \cite{Langberg_Effros} remains one of the key unresolved mysteries --- with direct connections to the edge-removal question \cite{Jalali_Effros_Ho} and  the existence of strong converses \cite{Kosut_Kliewer} in network information theory. Motivated by this curiosity, in this work we explore another important variant of PIR -- \emph{multiround} PIR (MPIR). Our  contributions are summarized next.

The classical PIR setting assumes that all the queries are simultaneously generated by the user. This assumption is also made in \cite{Sun_Jafar_PIR}. However, such a constraint is not essential to PIR. What if this constraint is relaxed, i.e., multiple rounds of queries and answers are allowed, such that the queries in each round of communication are generated by the user with the knowledge of the answers from all previous rounds? The resulting  variant of the PIR problem is the \emph{multiround} PIR (MPIR) problem (also known as interactive PIR \cite{Beimel_Ishai, Beimel_Ishai_Kushilevitz}). Multiround PIR has been noted as an intriguing possibility in several prior works \cite{PIRfirstjournal, Beimel_Ishai, Beimel_Ishai_Kushilevitz}. 
However, it  is not known whether there is any benefit of MPIR over single-round PIR.  Answering this question from a capacity perspective is the first contribution of this work. Specifically, we show that the capacity of MPIR is the same as the capacity of PIR, i.e., both are given by (\ref{eq:C}). Combined with previous results, this shows that there is no capacity advantage from multiround over single-round schemes, non-linear  over linear schemes or from $\epsilon$-error over zero-error schemes. Furthermore, we show that this is true even with $T$-privacy constraints.

To complement the capacity analysis, we consider another metric of interest --  storage overhead. Classical PIR assumes replicated databases, i.e., each database stores all the messages. For larger datasets, replication schemes incur substantial storage costs. Coding has been shown to be an effective way to reduce the storage costs in distributed data storage systems. Applications of coding to reduce the storage overhead for PIR have  attracted attention recently \cite{Shah_Rashmi_Kannan, Chan_Ho_Yamamoto, Fazeli_Vardy_Yaakobi, Tajeddine_Rouayheb, Rao_Vardy, Blackburn_Etzion, Blackburn_Etzion_Paterson, Banawan_Ulukus, Zhang_Wang_Wei_Ge, Wang_Skoglund}. In this context, our main contribution is an example ($N=2$ databases, $K=2$ messages) of a multiround, non-linear, $\epsilon$-error PIR scheme that achieves a strictly smaller storage overhead than the best  possible with a single-round, linear, zero-error scheme. The simplicity of  the scheme and the $N=K=2$ setting makes it an attractive point of reference for future work toward understanding the role of linear versus non-linear schemes, zero-error versus $\epsilon$-error  capacity, and single-round versus multiround communications. Interestingly, the scheme  reveals that coded storage is useful not only for reducing the storage overhead, but also it has a surprising benefit of enhancing the privacy of PIR.

\bigskip

{\it Notation: For  $n_1, n_2\in\mathbb{Z}, n_1\leq n_2$, define the notation $[n_1:n_2]$ as the set $\{n_1,n_1+1,\cdots, n_2\}$, $A(n_1:n_2)$ as the vector $(A(n_1), A(n_1+1),\cdots, A(n_2))$ and $A_{n_1:n_2}$ as the vector $(A_{n_1}, A_{n_1+1}, \cdots,A_{n_2})$. 
When $n_1 > n_2$, $[n_1:n_2]$ is a null set and $A(n_1:n_2), A_{n_1:n_2}$ are null vectors. For an index set $\mathcal{T} = \{i_1, i_2, \cdots, i_n\}$ such that $i_1<i_2<\cdots<i_n$, the notation $A_{\mathcal{T}}$ represents the vector $(A_{i_1}, A_{i_2}, \cdots, A_{i_n})$. 
The notation $X\sim Y$ is used to indicate that $X$ and $Y$ are identically distributed. }
\section{Problem Statement}\label{sec:model}
Let us start with a general problem statement that can then be specialized to various settings of interest. Consider $K$ independent messages $W_1, \cdots, W_K$, each comprised of $L$ i.i.d. uniform bits.
\begin{eqnarray}
&& H(W_1, \cdots, W_K) = H(W_1) + \cdots + H(W_K), \label{h1}\\
&& H(W_1) = \cdots = H(W_K) = L. \label{h2}
\end{eqnarray}
\noindent There are $N$ databases. Let $S_n$ denote the  information  that is stored at the $n^{th}$ database. 
\begin{eqnarray}
H(S_n|W_1, W_2, \cdots, W_K)&=&0, ~~\forall n\in[1:N]. \label{storagedet}
\end{eqnarray}
{\color{black} Define the storage overhead $\alpha$  as the ratio of the total amount of storage used by all databases  to the total amount of data. 
\begin{eqnarray}
\alpha&\define&\frac{\sum_{n=1}^NH(S_n)}{KL}.
\end{eqnarray}
}
For replication based schemes, each database stores all $K$ messages, so  $S_n=(W_1, W_2, \cdots, W_K)$, $H(S_n)=KL$, and the storage overhead, {\color{black}$\alpha = N$}.

A user privately generates $\theta$ uniformly from $[1:K]$ and wishes to retrieve $W_\theta$ while keeping $\theta$ a secret from each database. 


Prior works on capacity of PIR and its variants make certain (implicitly justified) assumptions of deterministic behavior, e.g., that the answers provided by the databases are deterministic functions of queries and messages. Here we will follow, instead, an explicit formulation. We allow randomness in the strategies followed by the user and the databases. This is accomplished by representing the actions of the user and the databases as functions of random variables. Let us use $\mathbb{F}$ to denote a random variable privately generated by the user, whose realization is not available to the databases.  Similarly, $\mathbb{G}$  is a random variable that determines the random strategies followed by the databases, and whose realizations are assumed to be known to all the databases and the user without loss of generality. $\mathbb{F}$ and $\mathbb{G}$ take values over the set of all deterministic strategies that the user or the databases can follow, respectively, associating each strategy with a certain probability. $\mathbb{F}$ and $\mathbb{G}$ are generated offline, i.e., before the realizations of the messages or the desired message index are known. Since these random variables are generated a-priori we must have
\begin{eqnarray}
&&H(\theta,\mathbb{F}, \mathbb{G}, W_1, \cdots, W_K) \notag \\
&=& H(\theta)+H(\mathbb{F}) + H(\mathbb{G}) +H(W_1) + \cdots + H(W_K) \label{maprandom}
\end{eqnarray}
The multiround PIR scheme proceeds as follows. Suppose $\theta=k$. In order to retrieve $W_k, k \in [1:K]$ privately, the user communicates with the databases over $\Gamma$ rounds. In the first round, the user privately generates $N$ random queries, $Q_1^{[k]}(1), Q_2^{[k]}(1), \cdots, Q_N^{[k]}(1)$. 
\begin{eqnarray}
H(Q_1^{[k]}(1), Q_2^{[k]}(1), \cdots, Q_N^{[k]}(1) | \mathbb{F}) = 0, ~~\forall k\in[1:K] \label{querydet}
\end{eqnarray}
The user  sends query $Q_n^{[k]}(1)$ to the $n^{th}$ database, $\forall n\in[1:N]$. Upon receiving $Q_n^{[k]}(1)$, the $n^{th}$ database generates an answering string $A_n^{[k]}(1)$. Without loss of generality, we assume that the answering string  is a function  of $Q_n^{[k]}(1)$, the stored information $S_n$, and the random variable $\mathbb{G}$.
\begin{eqnarray}
H(A_n^{[k]}(1) | Q_n^{[k]}(1), S_n, \mathbb{G}) = 0. \label{ansdet}
\end{eqnarray}
Each database returns to the user its answer $A_n^{[k]}(1)$. 

Proceeding similarly\footnote{One might wonder if the setting can be  further generalized by allowing sequential queries, i.e., allowing  the query to each database  to depend not only on the answers received from previous rounds, but also on the answers  received  from  other databases queried previously within the same round. We note that sequential queries are  already contained in our multiround framework, e.g., by querying only one database in each round (sending null queries to the remaining databases).}
, over the $\gamma^{th}$ round, $\gamma \in [2:\Gamma]$, the user generates $N$ queries $Q_1^{[k]}(\gamma), \cdots, Q_N^{[k]}(\gamma)$, which are  functions
of previous queries and answers and $\mathbb{F}$,
\begin{eqnarray}
H(Q_{1:N}^{[k]}(\gamma)| Q_{1:N}^{[k]}(1:\gamma-1), A_{1:N}^{[k]}(1:\gamma-1), \mathbb{F}) = 0 \label{querydet2}
\end{eqnarray}
The user sends query $Q_n^{[k]}(\gamma)$ to the $n^{th}$ database, which generates an answer $A_n^{[k]}(\gamma)$ and returns $A_n^{[k]}(\gamma)$ to the user. The answer is a  function  of all queries received so far, the stored information $S_n$, and $\mathbb{G}$,
\begin{eqnarray}
H(A_n^{[k]}(\gamma) | Q_n^{[k]}(1:\gamma), S_n, \mathbb{G}) = 0. \label{ansdetr}
\end{eqnarray}

At the end of $\Gamma$ rounds, from all the information that is now available to the user ($A_{1:N}^{[k]}(1:\Gamma), Q_{1:N}^{[k]}(1:\Gamma), \mathbb{F}$), the user  decodes the desired message $W_k$ according to a decoding rule that is specified by the PIR scheme. Let $P_e$ denote the probability of error  achieved with the specified decoding rule.

To protect the user's privacy, the $K$ possible values of the desired message index should be indistinguishable  from the perspective of any subset $\mathcal{T}\subset[1:N]$ of at most $T$ colluding databases, i.e., the following privacy constraint must be satisfied.
\begin{eqnarray}
[\mbox{$T$-Privacy}]~~(Q_{\mathcal{T}}^{[k]}(1:\Gamma), A_{\mathcal{T}}^{[k]}(1:\Gamma), \mathbb{G}, S_{\mathcal{T}}) &\sim&(Q_{\mathcal{T}}^{[k']}(1:\Gamma), A_{\mathcal{T}}^{[k']}(1:\Gamma), \mathbb{G}, S_{\mathcal{T}}) \label{tprivacy}
\\
&&\forall k,k'\in[1:K], \forall \mathcal{T}\subset [1:N], |\mathcal{T}|=T\nonumber
\end{eqnarray}

The PIR rate characterizes how many bits of desired information are retrieved per downloaded bit and is defined as follows.
\begin{eqnarray}
R = \frac{L}{D} \label{rate}
\end{eqnarray} 
where $D$ is the expected value\footnote{Alternatively, $D$ may be defined as the maximum download needed by the PIR scheme which (similar to choosing zero-error instead of $\epsilon$-error) weakens the converse and strengthens the achievability arguments in general. The capacity characterizations in this work, as well as previous works in \cite{Sun_Jafar_PIR, Sun_Jafar_TPIR,Sun_Jafar_SPIR} hold under either definition. This is because in every case, the upper bounds allow average download $D$, while the achievability only requires maximum download $D$.} of the total number of bits downloaded by the user from all the databases over all $\Gamma$ rounds. 

A rate $R_{\mbox{\footnotesize}}$ is said to be $\epsilon$-error achievable if there exists a sequence of PIR schemes, indexed by $L$, each of rate greater than or equal to $R_{\mbox{\footnotesize}}$, for which $P_e\rightarrow 0$ as $L\rightarrow\infty$. Note that for such a sequence of PIR schemes, from Fano's inequality we must have
\begin{eqnarray}
\mbox{[Correctness]} ~~~o(L) &=& \frac{1}{L}H(W_k | A_{1:N}^{[k]}(1:\Gamma), Q_{1:N}^{[k]}(1:\Gamma), \mathbb{F}) \notag\\
&\overset{(\ref{querydet})(\ref{querydet2})}{=}& \frac{1}{L} H(W_k | A_{1:N}^{[k]}(1:\Gamma), \mathbb{F}), ~~\forall k\in[1:K]\label{corr}
\end{eqnarray}
where $o(L)$ represents any term whose value approaches zero as $L$ approaches infinity. The supremum of $\epsilon$-error achievable rates is called the $\epsilon$-error capacity $C_{\epsilon \mbox{\footnotesize}}$. 

 A rate $R_{\mbox{\footnotesize}}$ is said to be zero-error achievable if there exists (for some $L$) a PIR scheme  of rate greater than or equal to $R_{\mbox{\footnotesize}}$ for which $P_e=0$. The supremum of zero-error achievable rates is called the zero-error capacity $C_{o \mbox{\footnotesize}}$. 
From the definitions, it is evident that 
\begin{eqnarray}
C_{o \mbox{\footnotesize}} \leq C_{\epsilon \mbox{\footnotesize}} 
\end{eqnarray}

\section{Results}\label{sec:main}
There are two main contributions in this work, summarized in the following sections.
\subsection{Capacity Perspective}\label{sec:cap}
We first consider the capacity benefits of multiple rounds of communication in the classical setting where each database stores all messages, i.e., storage is unconstrained.  We present our result in the general context of multiround PIR with $T$-privacy constraints (MTPIR). The MTPIR setting is obtained from the general problem statement by relaxing the storage overhead constraints, i.e., 
\begin{eqnarray*}
S_n&=&(W_1, W_2, \cdots, W_K), \forall n\in[1:N]\\
\alpha &=& N
\end{eqnarray*}
i.e., each database stores all the messages (replication). The following theorem presents the main result.
\begin{theorem}\label{thm:ts}
The capacity of MTPIR
$$ ~C_{o \mbox{\footnotesize}} = C_{\epsilon \mbox{\footnotesize}} = \left(1+T/N+T^2/N^2+\cdots+T^{K-1}/N^{K-1}\right)^{-1}.$$
\end{theorem}

The converse proof of Theorem \ref{thm:ts} is presented  in Section \ref{sec:tpir}. Achievability follows directly from \cite{Sun_Jafar_TPIR}. The following observations place the result in perspective.
\begin{enumerate}
\item The capacity of MTPIR matches the capacity of TPIR found in \cite{Sun_Jafar_TPIR}, i.e., multiple rounds do not increase capacity. 
\item Setting $T=1$ gives us the capacity of multiround PIR (MPIR) without $T$-privacy constraints. The capacity of MPIR matches the capacity of PIR found in \cite{Sun_Jafar_PIR}, i.e., multiple rounds do not increase capacity.
\item Since the achievability proofs in \cite{Sun_Jafar_TPIR, Sun_Jafar_PIR} only require linear and zero-error schemes, there is no capacity benefit of multiple rounds over single-round schemes, non-linear over linear schemes, or $\epsilon$-error over zero-error schemes.
\item For all $N, K, T, \Gamma$ the converse proof of Theorem \ref{thm:ts} generalizes the converse proofs of \cite{Sun_Jafar_TPIR, Sun_Jafar_PIR}. Remarkably, it requires only  Shannon information inequalities, i.e., sub-modularity of entropy.
\end{enumerate}

\subsection{Storage Overhead Perspective}
As summarized above, our first result shows that in a  broad sense -- with or without colluding databases -- there is no capacity benefit of multiple rounds over single-round communication, $\epsilon$-error over zero-error schemes or non-linear over linear schemes for PIR. This pessimistic finding may lead one to believe that there is little reason to further explore interactive communication, non-linear schemes or $\epsilon$-error schemes for PIR. As our main contribution in this section, we  offer an optimistic counterpoint by looking at the PIR problem from the perspective of storage overhead instead of capacity.  The counterpoint is made through a  counterexample. The counterexample is quite remarkable in itself as it shows from a storage overhead perspective not only the advantage of a multiround PIR scheme over all single-round PIR schemes, but also of a non-linear PIR scheme over all linear PIR schemes, and an $\epsilon$-error scheme over all zero-error schemes. 

For a counterexample the simplest setting is typically the most interesting. Therefore, in this section we will only consider the simplest non-trivial setting, with  $K=2$ messages,  $N=2$ databases, and $T=1$, i.e., no collusion  among databases. Recall that  for this setting the capacity is $C=2/3$. For our counterexample we explore the minimum storage overhead that is needed to achieve the rate $2/3$.

\begin{theorem}\label{thm:linear}
For $K = 2, N=2, T=1$, and for rate $2/3$,
\begin{enumerate}
\item there exists a  multiround, non-linear and $\epsilon$-error PIR scheme with storage overhead $$\alpha=3/4 + 3/8\log_2 3$$
which is less than $3/2$.
\item the  storage overhead of any  single-round, linear and zero-error PIR scheme is $$\alpha\geq 3/2$$ 
\end{enumerate}
\end{theorem}
The achievability arguments, including the multiround, non-linear and $\epsilon$-error PIR scheme that proves the first part of Theorem \ref{thm:linear} are presented in this section.  The proof of the second claim notably utilizes  Ingleton's inequality, which goes beyond submodularity, and is presented in  Section \ref{sec:linear}.

\subsubsection{A multiround, non-linear and $\epsilon$-error PIR scheme for $K=2, N=2, T=1$}
\label{sec:sw}


Define $w_1, w_2$ as two independent uniform binary random variables. Further,  define
\begin{eqnarray}
x_1 &=&w_1\land w_2\\
x_2 &=&(\sim w_1) \land (\sim w_2)\\
y_1 &=&w_1 \land  (\sim w_2) \\
y_2&=&(\sim w_1) \land  w_2 
\end{eqnarray}
where $\land$ and $\sim$ are the logical AND and NOT operators. 
Note the following,
\begin{eqnarray}
x_1=1&\Rightarrow&(w_1,w_2)=(1,1)\label{eq:11}\\
x_2=1&\Rightarrow&(w_1,w_2)=(0,0)\label{eq:00}\\
x_1=0&\Rightarrow&(w_1,w_2)=(y_1,y_2)\\
x_2=0&\Rightarrow&(w_1,w_2)=(\sim y_2,\sim y_1)\label{eq:10}
\end{eqnarray}

For ease of exposition, consider first the case where each message is only one bit long.  In this case, the  messages $W_1, W_2$, directly correspond to $w_1, w_2$, respectively.  
Denote the first database as DB1 and the second database as DB2. Regardless of whether the user desires $W_1$ or $W_2$, he flips a private fair coin, and requests the value of either $x_1$ or $x_2$ from DB1. If the answer is $1$, then according to (\ref{eq:11}) and (\ref{eq:00}) the user knows the values of both $w_1, w_2$ and no further information is requested from DB2. If the answer is $0$, then the user proceeds as follows.
\begin{itemize}
\item If $x_1 = 0$  and $W_1$ is desired, ask DB2 for  the value of $y_1$. Retrieve $w_1=y_1$.
\item If $x_1 = 0$ and  $W_2$ is desired, ask DB2 for the value of $y_2$. Retrieve $w_2=y_2$.
\item If $x_2 = 0$ and  $W_1$ is desired, ask DB2 for the value of $y_2$. Retrieve $w_1=\sim y_2$.
\item If $x_2 = 0$ and  $W_2$ is desired, ask DB2 for the value of $y_1$. Retrieve $w_2=\sim y_1$.
\end{itemize}
Note that in order to answer the user's queries, DB1 only needs to store $(x_1,x_2)$, and  DB2 only needs to store $(y_1,y_2)$. This observation is the key to not only the reduced storage overhead, but also the enhanced privacy of this scheme.

Further, in preparation for the proofs that follow, let us define another binary random variable $u$, which takes the value $u=0$ if no response is needed from DB2, and the value $u=1$ otherwise. Note that $u=0$ implies that $(y_1,y_2)=(0,0)$. On the other hand, if $u=1$, then $(y_1,y_2)$  takes the values $(0,0),(1,0), (0,1)$, each with probability $1/3$. Therefore, 
\begin{eqnarray}
H(y_1,y_2|u) &=&  1/4\times H(y_1,y_2|u=0) + 3/4 \times H(y_1,y_2 | u=1)\\
&=& 1/4 \times 0 + 3/4 \times H(1/3, 1/3, 1/3) = 3/4 \log_2 3
\end{eqnarray}

The correctness  of the scheme is obvious from (\ref{eq:11})-(\ref{eq:10}). Let us verify that the scheme is  private. Start with DB1. The query to DB1 is equally likely to be $x_1$ or $x_2$, regardless of the desired message index and the message realizations. Therefore, DB1 learns nothing about which message is retrieved. Next consider DB2. Let us prove that $(Q_2^{[1]}, y_1, y_2)\sim(Q_2^{[2]}, y_1, y_2)$. 
\begin{eqnarray*}
(\theta=1)~~~~~~~~~&&~~~~~~~~~(\theta=2)\\
\begin{array}{c|c}
{(Q_2^{[1]}, y_1, y_2)} & \mbox{Prob.}\\ \hline
(\emptyset, 0, 0) & 1/4 \\
(``y_1", 0, 0) & 1/8 \\
(``y_2", 0, 0) & 1/8 \\
(``y_1", 0, 1) & 1/8 \\
(``y_2", 0, 1) & 1/8 \\
(``y_1", 1, 0) & 1/8 \\
(``y_2", 1, 0) & 1/8 \\
\end{array}
&\sim&
\begin{array}{c|c}
{(Q_2^{[2]}, y_1, y_2)} & \mbox{Prob.}\\ \hline
(\emptyset, 0, 0) & 1/4 \\
(``y_1", 0, 0) & 1/8 \\
(``y_2", 0, 0) & 1/8 \\
(``y_1", 0, 1) & 1/8 \\
(``y_2", 0, 1) & 1/8 \\
(``y_1", 1, 0) & 1/8 \\
(``y_2", 1, 0) & 1/8 \\
\end{array}
\end{eqnarray*}
where the double quote notation  around a random variable represents the query about its realization. The computation of the joint distribution values is straightforward. We present the derivation here for one case. All other cases follow similarly. From the law of total probability, we have
\begin{align}
\lefteqn{\Pr\Big( (Q_2^{[1]}, y_1, y_2) = (``y_1", 0, 1) \Big)} \notag\\
&= \Pr\Big( (Q_2^{[1]}, y_1, y_2) = (``y_1", 0, 1) | (Q_1^{[1]}, w_1, w_2)  = (``x_1", 0,1) \Big)  \times \Pr\Big(  (Q_1^{[1]}, w_1, w_2)  = (``x_1", 0,1)  \Big) \notag\\
& + \Pr\Big( (Q_2^{[1]}, y_1, y_2) = (``y_1", 0, 1) | (Q_1^{[1]}, w_1, w_2)  = (``x_2", 0,1) \Big)  \times \Pr\Big( (Q_1^{[1]}, w_1, w_2)  = (``x_2", 0,1)  \Big) \label{zero} \\
&= 1\times 1/8 + 0\times 1/8 = 1/8
\end{align}
Similarly,
\begin{align}
\lefteqn{\Pr\Big( (Q_2^{[2]}, y_1, y_2) = (``y_1", 0, 1) \Big)} \notag\\
&= \Pr\Big( (Q_2^{[2]}, y_1, y_2) = (``y_1", 0, 1) | (Q_1^{[2]}, w_1, w_2)  = (``x_1", 0,1) \Big)  \times \Pr\Big(  (Q_1^{[2]}, w_1, w_2)  = (``x_1", 0,1)  \Big) \notag\\
& + \Pr\Big( (Q_2^{[2]}, y_1, y_2) = (``y_1", 0, 1) | (Q_1^{[2]}, w_1, w_2)  = (``x_2", 0,1) \Big)  \times \Pr\Big( (Q_1^{[2]}, w_1, w_2)  = (``x_2", 0,1)  \Big) \label{zero} \\
&= 0\times 1/8 + 1\times 1/8 = 1/8
\end{align}
Thus, $\Pr\Big( (Q_2^{[1]}, y_1, y_2) = (``y_1", 0, 1)\Big)=\Pr\Big( (Q_2^{[1]}, y_1, y_2) = (``y_1", 0, 1)\Big)$. All other cases are verified similarly. Then, since the  distribution of $(Q_2^{[\theta]}, y_1, y_2)$ does not depend on $\theta$, and the answers are only deterministic functions of the query and the stored information, it follows that the scheme is private.

Next consider the $L$ length extension of this PIR scheme, where each desired bit is retrieved independently as described above. Under the $L$ length extension, $W_1, W_2, X_1, X_2, Y_1, Y_2,U$ are sequences of length $L$, such that the sequence of tuples $[(W_1(l), W_2(l)$,  $X_1(l), X_2(l)$, $Y_1(l), Y_2(l),U(l))]_{l=1}^L$ is  i.i.d. $\sim(w_1,w_2,x_1,x_2,y_1,y_2,u)$. Since the  extension is obtained by repeated independent applications  of the PIR scheme described above for retrieving each message bit, it follows trivially that the extended PIR scheme is also correct and private. The  purpose for the $L$ length extension, with $L\rightarrow\infty$, is to invoke fundamental limits of data compression which optimize both the data rates and the storage overhead as explained next.

Let us  show that the rate   $2/3$ is achieved asymptotically as $L\rightarrow\infty$. 
We take advantage of the fact that  the answers from the databases are not uniformly distributed, and therefore the sequence of answers from each database is compressible. With optimal compression, the user downloads $H(1/4, 3/4)$ bits per desired message bit from DB1. This is because, for each retrieved bit, the answer from DB1 takes the value $1$ with probability $1/4$ and $0$ with probability $3/4$. From DB2, we download $1/4 \times 0 + 3/4 \times H(1/3, 2/3) = 3/4 H(1/3, 2/3)$ bits per desired message bit, because with probability  $1/4$ (when the answer from DB1 is $1$), no response is requested from DB2 and otherwise within the remaining space of probability measure $3/4$  (when the answer from DB1 is $0$), the answer from DB2 is $1$ with conditional probability $1/3$ and $0$ with conditional probability $2/3$. Therefore the total download is $H(1/4, 3/4) + 3/4 H(1/3, 2/3) = 3/2$ bits per desired message bit and the rate achieved is $2/3$.

Next let us determine the storage requirements of this scheme. DB1 needs $(X_1, X_2)$ to answer the user's queries, so with optimal compression, it needs to store $H(x_1,x_2)=H(1/4, 1/4, 1/2) = 3/2$ bits per desired message bit. One might naively imagine that the same storage requirement also applies to DB2, because DB2 similarly needs the values $(Y_1, Y_2)$ to answer the user's queries. However, this is not true, because  the query sent to DB2 already contains some information about the message realizations,\footnote{Note that the query sent to DB2 is independent of the desired message index but not the message realizations.} and this \emph{side-information} allows DB2 to reduce its storage requirement by taking advantage of Slepian Wolf coding \cite{Slepian_Wolf, Cover_Thomas} (distributed compression with decoder side information).

The key  is to realize that DB2 does not need to know $(Y_1,Y_2)$ until after it receives the query from the user. The query from the user includes $U$ as side information. Therefore, using Slepian Wolf coding, DB2 is able to optimally compress  the i.i.d. sequence $(Y_1, Y_2)$  to the conditional entropy $H(y_1,y_2|u)$ bits per desired message bit and still decode the $(Y_1,Y_2)$ sequence when it is needed, i.e., after the query is provided by the user. Thus, the total storage required by this PIR scheme is $3/2+3/4\log_23$ bits per bit of desired message. Since there are two messages, the storage overhead is $3/4+3/8 \log_2 3$.

The following observations are useful to place the new PIR scheme in perspective.
\begin{enumerate}
\item The optimal compression guarantees are only available in the $\epsilon$-error sense. Therefore, this PIR scheme is  essentially an $\epsilon$-error scheme. 
\item The multiround scheme is in fact a sequential PIR scheme that utilizes only one round of queries for each database (two rounds total since there are two databases). 
\item The scheme is  essentially non-linear because, e.g., the logical AND operator is non-linear. 
\item Since the multiround, non-linear and $\epsilon$-error aspects are all essential for \emph{this} scheme to create an advantage in terms of storage overhead,  it is an intriguing question whether all three aspects are necessary in \emph{general} or if it is  possible to achieve storage overhead less than $3/2$ through another scheme while sacrificing at least one of the three aspects.
\item  A key insight from this PIR scheme is the surprising privacy benefit of storage overhead optimization. By not storing all the information at each database, and by optimally compressing the stored information, not only do we reduce the storage overhead, but also we enable stronger privacy guarantees than would hold otherwise. Note that if each database stores all the information (both $W_1$ and $W_2$), then the scheme is not private. To see this, suppose $(w_1,w_2)=(1,1)$. This would be known to DB2 because it stores both messages. Under this circumstance, DB2  knows that if the user asks for $y_2$, then his desired message must be $W_1$ and if the user asks for $y_1$ then his desired message must be $W_2$. Thus, storing all the information at each database would result in loss of privacy. As another example, we note that if the data is not in its optimally compressed form, i.e., $w_1$ and/or $w_2$ are not uniformly distributed then again the PIR scheme would lose privacy. To see this, suppose Pr($w_1=1$)=Pr$(w_2=1) > 1/2$. Then  DB2 is more likely to be asked for $y_1$ if the desired message is $W_2$ than if the desired message is $W_1$. On the other hand, note that   optimal data compression is a pre-requisite in any case for the optimization of rate and storage overhead.\footnote{Since optimal compression limits are typically achieved asymptotically, if the data is not assumed to be uniform a-priori, then as noted by \cite{Beimel_Ishai, Beimel_Ishai_Kushilevitz} the privacy guarantees would also be subject to  $\epsilon$-leakage that approaches zero as message length approaches infinity.} 

\item Let us consider momentarily the restricted message size setting, where each message is only $L=1$ bit long. Then it is easy to see that any single-round scheme (all queries generated simultaneously) must download at least $2$ bits on average, but our multiround scheme requires an expected download of only $1+3/4=7/4$ bits. Thus, even though the download advantage of multiround PIR disappears under unconstrained message lengths, for constrained message lengths there are benefits of multiround PIR.
\end{enumerate}

\subsubsection{A single-round, linear and zero-error scheme for $K=2, N=2, T=1$}
For comparison, the corresponding scheme from \cite{Sun_Jafar_PIR} which also achieves rate $2/3$ is reproduced below. This will be shown to be the optimal  single-round, linear, zero-error scheme for storage overhead in Section \ref{sec:linear}.  Denote the messages, each comprised of $4$ bits,  as $W_1 = (a_1, a_2, a_3, a_4), W_2 = (b_1, b_2, b_3, b_4)$. The downloaded information from each database is shown below.
\begin{eqnarray*}
\centering
\renewcommand*{\arraystretch}{1}
\begin{array}{|c|c|c|c|c|}\hline
&\multicolumn{2}{c}{\mbox{Prob. $1/2$}} \vline&\multicolumn{2}{c}{\mbox{Prob. $1/2$}} \vline\\ \cline{2-5}
    & \mbox{Want $W_1$}  &\mbox{Want $W_2 $}& \mbox{Want $W_1$} &\mbox{Want $W_2$} \\
   \hline		
\mbox{Database $1$}&a_1, b_1, a_2 + b_2 &a_1, b_1, a_2 + b_2& a_3, b_3, a_4 + b_4&a_3, b_3, a_4 + b_4\\ \hline
\mbox{Database $2$}& a_4, b_2, a_3 + b_1 &a_2, b_4, a_1 + b_3&a_2, b_4, a_1 + b_3&a_4, b_2, a_3 +b_1\\ \hline
\end{array}
\end{eqnarray*}
\noindent The scheme achieves rate $2/3$ and is linear, single-round, and zero-error. 
A total of $6$ bits are stored at each database
\begin{eqnarray}
S_1 = (a_1, a_3, b_1, b_3, a_2+b_2, a_4+b_4)\\
S_2 = (a_2, a_4, b_2, b_4, a_3+b_1, a_1+b_3)
\end{eqnarray} 
Thus, the storage overhead  is $3/2$.

\section{Proof of  Theorem \ref{thm:ts}}\label{sec:tpir}
We first present two useful lemmas. Note that in the proofs, the relevant equations needed to justify each step are specified by the equation numbers set on top of the (in)equality symbols.
\begin{lemma}\label{lemma:tinduction}
For all $k \in [2:K]$,
\begin{eqnarray}
\lefteqn{I(W_{k:K}; Q_{1:N}^{[k-1]}(1:\Gamma), A_{1:N}^{[k-1]}(1:\Gamma), \mathbb{F}|W_{1:k-1}, \mathbb{G})} \notag\\
&\geq& \frac{T}{N} I(W_{k+1:K}; Q_{1:N}^{[k]}(1:\Gamma), A_{1:N}^{[k]}(1:\Gamma), \mathbb{F} |W_{1:k}, \mathbb{G}) + \frac{LT}{N}(1-o(L))\label{eq:tinduction}.
\end{eqnarray}
\end{lemma}

{\it Proof:} 
\begin{eqnarray}
\lefteqn{N I(W_{k:K}; Q_{1:N}^{[k-1]}(1:\Gamma), A_{1:N}^{[k-1]}(1:\Gamma), \mathbb{F}|W_{1:k-1}, \mathbb{G}) }\notag\\
&\geq& \frac{N}{\binom{N}{T}} \sum_{\mathcal{T} {\subset} [1:N], |\mathcal{T}| = T} I(W_{k:K}; Q_{\mathcal{T}}^{[k-1]}(1:\Gamma), A_{\mathcal{T}}^{[k-1]}(1:\Gamma)|W_{1:k-1}, \mathbb{G}) \label{l11}\\
&\overset{(\ref{tprivacy})}{=}& \frac{N}{\binom{N}{T}}  \sum_{\mathcal{T} \subset [1:N], |\mathcal{T}| = T} I(W_{k:K}; Q_{\mathcal{T}}^{[k]}(1:\Gamma), A_{\mathcal{T}}^{[k]}(1:\Gamma)|W_{1:k-1}, \mathbb{G}) \\
&\overset{}{=}& \frac{N}{\binom{N}{T}} \sum_{\mathcal{T} \subset [1:N], |\mathcal{T}| = T} \sum_{\gamma=1}^\Gamma I(W_{k:K}; Q_{\mathcal{T}}^{[k]}(\gamma), A_{\mathcal{T}}^{[k]}(\gamma)|Q_{\mathcal{T}}^{[k]}(1:\gamma-1), A_{\mathcal{T}}^{[k]}(1:\gamma-1), W_{1:k-1}, \mathbb{G}) \notag\\
&\overset{}{\geq}& \frac{N}{\binom{N}{T}} \sum_{\mathcal{T} \subset [1:N], |\mathcal{T}| = T} \sum_{\gamma=1}^\Gamma I(W_{k:K}; A_{\mathcal{T}}^{[k]}(\gamma)|Q_{\mathcal{T}}^{[k]}(1:\gamma), A_{\mathcal{T}}^{[k]}(1:\gamma-1), W_{1:k-1}, \mathbb{G}) \\
&\overset{(\ref{ansdet})(\ref{ansdetr})}{=}& \frac{N}{\binom{N}{T}} \sum_{\mathcal{T} \subset [1:N], |\mathcal{T}| = T} \sum_{\gamma=1}^\Gamma H(A_{\mathcal{T}}^{[k]}(\gamma)|Q_{\mathcal{T}}^{[k]}(1:\gamma), A_{\mathcal{T}}^{[k]}(1:\gamma-1), W_{1:k-1}, \mathbb{G}) \\
&\overset{}{\geq}& \frac{N}{\binom{N}{T}} \sum_{\mathcal{T} \subset [1:N], |\mathcal{T}| = T} \sum_{\gamma=1}^\Gamma H(A_{\mathcal{T}}^{[k]}(\gamma)|Q_{1:N}^{[k]}(1:\gamma), A_{1:N}^{[k]}(1:\gamma-1), W_{1:k-1}, \mathbb{F}, \mathbb{G}) \\
&\overset{}{\geq}& T \sum_{\gamma=1}^\Gamma H(A_{1:N}^{[k]}(\gamma)|Q_{1:N}^{[k]}(1:\gamma), A_{1:N}^{[k]}(1:\gamma-1), W_{1:k-1}, \mathbb{F}, \mathbb{G}) ~~\mbox{\footnotesize (Han's inequality \cite{Cover_Thomas})}\\
&\overset{(\ref{ansdet})(\ref{ansdetr})}{=}& T \sum_{\gamma=1}^\Gamma I(W_{k:K}; A_{1:N}^{[k]}(\gamma)|Q_{1:N}^{[k]}(1:\gamma), A_{1:N}^{[k]}(1:\gamma-1), W_{1:k-1}, \mathbb{F}, \mathbb{G})  \\
&\overset{(\ref{querydet})(\ref{querydet2})}{=}& T \sum_{\gamma=1}^\Gamma I(W_{k:K}; Q_{1:N}^{[k]}(\gamma), A_{1:N}^{[k]}(\gamma)|Q_{1:N}^{[k]}(1:\gamma-1), A_{1:N}^{[k]}(1:\gamma-1), W_{1:k-1}, \mathbb{F}, \mathbb{G}) \\
&=&TI(W_{k:K}; Q_{1:N}^{[k]}(1:\Gamma), A_{1:N}^{[k]}(1:\Gamma)|W_{1:k-1},  \mathbb{F},\mathbb{G})\\
&\overset{(\ref{corr})}{=}&  TI(W_{k:K}; W_k, Q_{1:N}^{[k]}(1:\Gamma), A_{1:N}^{[k]}(1:\Gamma) |W_{1:k-1}, \mathbb{F}, \mathbb{G}) - o(L)LT\\
&\overset{}{=}& T I(W_{k:K}; W_k|W_{1:k-1}, \mathbb{F}, \mathbb{G}) - o(L)LT \notag\\
&& +~ {\color{black}T}I(W_{k+1:K}; Q_{1:N}^{[k]}(1:\Gamma), A_{1:N}^{[k]}(1:\Gamma) |W_{1:k}, \mathbb{F}, \mathbb{G})\\
&\overset{(\ref{maprandom})}{=}& LT(1-o(L)) +{\color{black}T} I(W_{k+1:K}; Q_{1:N}^{[k]}(1:\Gamma), A_{1:N}^{[k]}(1:\Gamma) |W_{1:k}, \mathbb{F}, \mathbb{G})\\
&\overset{(\ref{maprandom})}{=}& LT(1-o(L)) +{\color{black}T} I(W_{k+1:K}; Q_{1:N}^{[k]}(1:\Gamma), A_{1:N}^{[k]}(1:\Gamma),\mathbb{F} |W_{1:k}, \mathbb{G})
\end{eqnarray}
\hfill\QED

\begin{lemma}\label{lemma:int}
\begin{align}
I(W_{2:K}; Q_{1:N}^{[1]}(1:\Gamma), A_{1:N}^{[1]}(1:\Gamma), \mathbb{F}|W_1, \mathbb{G}) \leq L(1/R_{}  - 1 + o(L))
. \label{eq:int}
\end{align}
\end{lemma}

{\it Proof: }
\begin{eqnarray}
\lefteqn{ I(W_{2:K}; Q_{1:N}^{[1]}(1:\Gamma), A_{1:N}^{[1]}(1:\Gamma),  \mathbb{F}|W_1, \mathbb{G})} \notag\\
&\overset{(\ref{maprandom})}{=}& I(W_{2:K}; Q_{1:N}^{[1]}(1:\Gamma), A_{1:N}^{[1]}(1:\Gamma), W_1, \mathbb{F}, \mathbb{G}) \\
&\overset{(\ref{querydet})(\ref{querydet2})}{=}& I(W_{2:K}; A_{1:N}^{[1]}(1:\Gamma), W_1,\mathbb{F}, \mathbb{G}) \label{l21}\\
&\overset{}{=}& I(W_{2:K}; A_{1:N}^{[1]}(1:\Gamma), \mathbb{F}, \mathbb{G}) + I(W_{2:K};W_1|A_{1:N}^{[1]}(1:\Gamma), \mathbb{F}, \mathbb{G})\\
&\overset{(\ref{maprandom})(\ref{corr})}{=}& I(W_{2:K}; A_{1:N}^{[1]}(1:\Gamma) | \mathbb{F}, \mathbb{G}) + o(L)L \\
&\overset{}{=}& H(A_{1:N}^{[1]}(1:\Gamma) | \mathbb{F}, \mathbb{G}) - H(A_{1:N}^{[1]}(1:\Gamma) | \mathbb{F}, \mathbb{G}, W_{2:K}) + o(L)L \\
&\overset{(\ref{rate})}{\leq}& L/R_{}   - H(A_{1:N}^{[1]}(1:\Gamma) | \mathbb{F}, \mathbb{G}, W_{2:K}) + o(L)L \\
&\overset{(\ref{corr})}{=}& L/R_{}   - H(W_1,A_{1:N}^{[1]}(1:\Gamma) | \mathbb{F}, \mathbb{G}, W_{2:K})  + o(L)L\\
&\overset{}{\leq}& L/R_{}  - H(W_1| \mathbb{F}, \mathbb{G}, W_{2:K}) + o(L)L\\
&\overset{(\ref{maprandom})}{=}& L/R_{}  - L + o(L)L= L(1/R_{}  - 1 + o(L)) \label{eq:final}
\end{eqnarray}
\hfill\QED

\noindent With Lemma \ref{lemma:tinduction} and Lemma \ref{lemma:int}, we are ready to prove the converse.

\subsection*{Rate Outerbound}
Starting from $k = 2$ and applying (\ref{eq:tinduction}) repeatedly for $k \in [3:K]$,
\begin{eqnarray}
\lefteqn{I(W_{2:K}; Q_{1:N}^{[1]}(1:\Gamma), A_{1:N}^{[1]}(1:\Gamma), \mathbb{F}|W_1, \mathbb{G}) }\notag\\
&\overset{(\ref{eq:tinduction})}{\geq}& \frac{T}{N} I(W_{3:K}; Q_{1:N}^{[2]}(1:\Gamma), A_{1:N}^{[2]}(1:\Gamma), \mathbb{F} |W_1, W_2, \mathbb{G}) + \frac{LT(1-o(L))}{N} \notag\\
&\overset{(\ref{eq:tinduction})}{\geq}& \cdots \\
&\overset{(\ref{eq:tinduction})}{\geq}& \frac{T^{K-2}}{N^{K-2}} I(W_{K}; Q_{1:N}^{[K-1]}(1:\Gamma), A_{1:N}^{[K-1]}(1:\Gamma),\mathbb{F} |W_{1:K-1}, \mathbb{G})\notag\\
&&+~ \frac{LT(1-o(L))}{N}  + \cdots + \frac{LT^{K-2}(1-o(L))}{N^{K-2}} \notag\\
&\overset{(\ref{eq:tinduction})}{\geq}& \frac{T^{K-2}}{N^{K-2}} \frac{LT(1-o(L))}{N} + \frac{LT(1-o(L))}{N}  + \cdots + \frac{LT^{K-2}(1-o(L))}{N^{K-2}} \\
&=& L(1-o(L))(T/N+\cdots+T^{K-1}/N^{K-1}) \label{eq:ind_final}
\end{eqnarray}
Combining (\ref{eq:ind_final}) and (\ref{eq:int}), we have
\begin{eqnarray}
L(1/R_{}  - 1 + o(L)) \geq L(1-o(L))(T/N+\cdots+T^{K-1}/N^{K-1})
\end{eqnarray}
Normalizing by $L$ and letting $L$ go to infinity gives us
\begin{eqnarray}
1/R_{}  - 1 &\geq& T/N+\cdots+T^{K-1}/N^{K-1} \\
\Rightarrow R_{} &\leq& (1+T/N+\cdots+T^{K-1}/N^{K-1})^{-1}
\end{eqnarray}
thus, the proof is complete.

\section{Proof of Theorem \ref{thm:linear} -- Statement 2.}\label{sec:linear}
We show that when $K =2, N =2, T=1, \Gamma=1$ and the rate equals 2/3, the storage overhead of all zero-error, linear, and single-round PIR schemes is no less than $3/2$. 
Since we only consider single-round schemes in this section, we will simplify the notation, e.g., instead of $Q_2^{[1]}(1)$ we write simply $Q_2^{[1]}$.  In addition, without loss of generality, let us make the following simplifying assumptions.
\begin{enumerate}
\item We assume that the PIR scheme is symmetric, in that 
\begin{eqnarray}
H(A_1^{[1]}|\mathbb{F}, \mathbb{G}) &=& H(A_2^{[1]}| \mathbb{F}, \mathbb{G}) = H(A_2^{[2]}| \mathbb{F}, \mathbb{G}) \label{eq:caps}\\
H(S_1) &=& H(S_2) \label{s_sym}
\end{eqnarray}
Given any (asymmetric) PIR scheme that retrieves messages of size $L$, a symmetric PIR scheme with the same rate and storage overhead that retrieves messages of size $NL$ is obtained by repeating the original scheme $N$ times, and in the $n^{th}$ repetition shifting the database indices cyclically by $n$. This symmetrization process is described  in Theorem \ref{thm:symmetry} (see Section \ref{sec:symmetry}).

\item We  assume that $Q_1^{[1]} = Q_1^{[2]}$, i.e., the query for the first database is chosen without the knowledge of the desired message index. There is no loss of generality in this assumption because of the privacy constraint, which requires that $Q_1^{[\theta]}$ is independent of $\theta$.\footnote{Note that instead of $Q_1^{[1]} = Q_1^{[2]}$, we could equivalently assume that $Q_2^{[1]} = Q_2^{[2]}$ without of loss of generality (because privacy also requires that $Q_2^{[\theta]}$ is  independent of $\theta$). However, if we simultaneously assume both  $Q_1^{[1]} = Q_1^{[2]}$ \emph{and} $Q_2^{[1]} = Q_2^{[2]}$, then there is a loss of generality because together $(Q_1^{[\theta]},Q_2^{[\theta]})$ is \emph{not} required to be independent of $\theta$ by the privacy constraint.} Note that this also means that $A_1^{[1]} = A_1^{[2]}$. 
\end{enumerate}

Our goal is to prove a lower bound on the storage overhead. Since the PIR scheme is symmetric by assumption, the storage overhead is $(H(S_1)+H(S_2))/2L=H(S_2)/L$. Furthermore,  $H(S_2) \geq H(A_2^{[1]}, {A}_2^{[2]}|\mathbb{F},\mathbb{G})$, so we will prove a lower bound on $H(A_2^{[1]}, {A}_2^{[2]}|\mathbb{F},\mathbb{G})$ instead.

Let us start with a useful lemma that holds for all linear and non-linear schemes. 
\begin{lemma}\label{lemma:id}
\begin{eqnarray}
H(A_1^{[1]}|W_1, \mathbb{F}, \mathbb{G}) &=& H({A}_2^{[2]}|W_1, \mathbb{F}, \mathbb{G}) = H({A}_2^{[2]}|W_2,\mathbb{F}, \mathbb{G})  = L/2 \label{l2} \\
H(A_2^{[2]}|W_1, A_2^{[1]}, \mathbb{F}, \mathbb{G}) &=& H(A_2^{[2]}|W_2, A_2^{[1]}, \mathbb{F}, \mathbb{G}) = L/2  \label{a12}
\end{eqnarray}
\end{lemma}
{\it Proof:} We prove (\ref{l2}) first. On the one hand, after substituting\footnote{Since we are considering only zero-error schemes, the $o(L)$ term in Lemma \ref{lemma:int} is exactly $0$.} $R=2/3$ in Lemma \ref{lemma:int}, from (\ref{l21})  we have
\begin{eqnarray}
L/2 &\geq& I(W_{2}; A_{1}^{[1]}, A_{2}^{[1]}, W_{1}, \mathbb{F}, \mathbb{G}) \\
&\overset{(\ref{maprandom})}{=}&I(W_{2}; A_{1}^{[1]}, A_{2}^{[1]}|W_{1}, \mathbb{F}, \mathbb{G}) \\
&\overset{(\ref{querydet})(\ref{ansdet})(\ref{storagedet})}{=}& H(A_{1}^{[1]}, A_{2}^{[1]}|W_{1}, \mathbb{F}, \mathbb{G}) \label{eq:by}\\
\Rightarrow L/2 &\overset{}{\geq}& H(A_{1}^{[1]}|W_{1}, \mathbb{F}, \mathbb{G}) \label{aa1}\\
 \mbox{and } L/2 &\overset{}{\geq}& H(A_{2}^{[1]}|W_{1}, \mathbb{F}, \mathbb{G}) \label{aa2}
\end{eqnarray}
On the other hand, from (\ref{l11}) in Lemma \ref{lemma:tinduction}, we have
\begin{eqnarray}
L &\leq& I(W_{2}; Q_{1}^{[1]}, A_{1}^{[1]}|W_{1},\mathbb{G}) + I(W_{2}; Q_{2}^{[1]}, A_{2}^{[1]}|W_{1},\mathbb{G}) \label{dir}\\
&\leq& I(W_{2}; Q_{1}^{[1]}, A_{1}^{[1]}, \mathbb{F}|W_{1},\mathbb{G}) + I(W_{2}; Q_{2}^{[1]}, A_{2}^{[1]}, \mathbb{F}|W_{1},\mathbb{G})\\
&\overset{(\ref{maprandom})}{=}& I(W_{2}; Q_{1}^{[1]}, A_{1}^{[1]}|W_{1}, \mathbb{F}, \mathbb{G}) + I(W_{2}; Q_{2}^{[1]}, A_{2}^{[1]}|W_{1}, \mathbb{F},\mathbb{G})\\
&\overset{(\ref{querydet})(\ref{ansdet})(\ref{storagedet})}{=}&  H(A_{1}^{[1]}|W_{1},\mathbb{F}, \mathbb{G}) + H(A_{2}^{[1]}|W_{1},\mathbb{F},\mathbb{G}) \label{aa}
\end{eqnarray}

Combining (\ref{aa1}), (\ref{aa2}) and (\ref{aa}), we have shown that 
\begin{eqnarray}
H(A_1^{[1]}|W_1, \mathbb{F}, \mathbb{G}) = H(A_{2}^{[1]}|W_{1},\mathbb{F},\mathbb{G}) = L/2 \label{a21}
\end{eqnarray} 
Symmetrically,  it follows that $H({A}_2^{[2]}|W_2,\mathbb{F}, \mathbb{G})  = L/2$.
We are left to prove $H({A}_2^{[2]}|W_1, \mathbb{F}, \mathbb{G}) = L/2$. On the one hand, from (\ref{aa1}) and (\ref{aa2}), we have
{\color{black} 
\begin{eqnarray}
L/2 &\overset{}{\geq}& H(A_{1}^{[1]}|W_{1}, \mathbb{F}, \mathbb{G}) \overset{}{=} H(A_{1}^{[2]}|W_{1}, \mathbb{F}, \mathbb{G}) ~~~(\mbox{Using}~A_1^{[1]} = A_1^{[2]}) \label{aa3}\\
 L/2 &\overset{}{\geq}& H(A_{2}^{[1]}|W_{1}, \mathbb{F}, \mathbb{G}) \\
 &\overset{(\ref{querydet})}{=}& H(A_{2}^{[1]}|W_{1}, Q_{2}^{[1]}, \mathbb{F}, \mathbb{G}) \\
 &\overset{}{=}& H(A_{2}^{[1]}|W_{1}, Q_{2}^{[1]}, \mathbb{G}) \label{m1}\\
 &\overset{}{=}& H(A_{2}^{[2]}|W_{1}, Q_{2}^{[2]}, \mathbb{G}) \label{ee1} \\
 &\overset{}{=}& H(A_{2}^{[2]}|W_{1}, Q_{2}^{[2]}, \mathbb{F}, \mathbb{G}) \label{m2}\\
&\overset{(\ref{querydet})}{=}& H(A_{2}^{[2]}|W_{1}, \mathbb{F}, \mathbb{G}) \label{aa4} 
\end{eqnarray}
where (\ref{ee1}) follows from the fact that for single-round PIR, the desired message index is independent of the messages, queries and answers, i.e., from (\ref{maprandom}), we have
\begin{eqnarray}
&& I(\theta; W_1, W_2, \mathbb{F}, \mathbb{G}) = 0 \\
&\overset{(\ref{querydet})}{\Longrightarrow}&  I(\theta; W_1, W_2, \mathbb{F}, \mathbb{G}, Q_{2}^{[\theta]}) = 0 \\
&\overset{(\ref{ansdet})(\ref{storagedet})}{\Longrightarrow}& I(\theta; W_1, W_2, \mathbb{F}, \mathbb{G}, Q_{2}^{[\theta]}, A_{2}^{[\theta]}) = 0 \\
&\overset{}{\Longrightarrow}& A_{2}^{[1]}, W_1, Q_{2}^{[1]}, \mathbb{G}  \sim A_{2}^{[2]}, W_1, Q_{2}^{[2]}, \mathbb{G}
\end{eqnarray}
(\ref{m1}) and (\ref{m2}) are due to the Markov chain $\mathbb{F} - (W_1, Q_2^{[k]},\mathbb{G}) - A_2^{[k]}, k = 1,2$, which is proved as follows.
\begin{eqnarray}
I(A_2^{[k]}; \mathbb{F} | W_1, Q_2^{[k]}, \mathbb{G}) &\leq& I(A_2^{[k]}, S_2; \mathbb{F} | W_1, Q_2^{[k]}, \mathbb{G})\\
&=& I(S_2; \mathbb{F} | W_1, Q_2^{[k]}, \mathbb{G}) + I(A_2^{[k]}; \mathbb{F} | W_1, Q_2^{[k]}, \mathbb{G}, S_2)\\
&\overset{(\ref{ansdet})}{=}& I(S_2; \mathbb{F} | W_1, Q_2^{[k]}, \mathbb{G}) \\
&\overset{}{\leq}& I(S_2, W_2; \mathbb{F} | W_1, Q_2^{[k]}, \mathbb{G}) \\
&\overset{}{=}& I(W_2; \mathbb{F} | W_1, Q_2^{[k]}, \mathbb{G}) + I(S_2; \mathbb{F} | Q_2^{[k]}, \mathbb{G},W_1, W_2)\\
&\overset{(\ref{storagedet})}{\leq}& I(W_2; \mathbb{F}, W_1, Q_2^{[k]}, \mathbb{G}) \\
&\overset{(\ref{querydet})(\ref{maprandom})}{=}& 0 
\end{eqnarray}


}

\noindent On the other hand, from (\ref{dir}), we have
\begin{eqnarray}
L &\leq& I(W_{2}; Q_{1}^{[1]}, A_{1}^{[1]}|W_{1},\mathbb{G}) + I(W_{2}; Q_{2}^{[1]}, A_{2}^{[1]}|W_{1},\mathbb{G}) \\
&\overset{(\ref{tprivacy})}{=}& I(W_{2}; Q_{1}^{[2]}, A_{1}^{[2]}|W_{1},\mathbb{G}) + I(W_{2}; Q_{2}^{[2]}, A_{2}^{[2]}|W_{1},\mathbb{G}) \\
&\leq& I(W_{2}; Q_{1}^{[2]}, A_{1}^{[2]}, \mathbb{F}|W_{1},\mathbb{G}) + I(W_{2}; Q_{2}^{[2]}, A_{2}^{[2]}, \mathbb{F}|W_{1},\mathbb{G})\\
&\overset{(\ref{maprandom})}{=}& I(W_{2}; Q_{1}^{[2]}, A_{1}^{[2]}|W_{1}, \mathbb{F}, \mathbb{G}) + I(W_{2}; Q_{2}^{[2]}, A_{2}^{[2]}|W_{1}, \mathbb{F},\mathbb{G})\\
&\overset{(\ref{querydet})(\ref{ansdet})(\ref{storagedet})}{=}&  H(A_{1}^{[2]}|W_{1},\mathbb{F}, \mathbb{G}) + H(A_{2}^{[2]}|W_{1},\mathbb{F},\mathbb{G}) \label{aaa}
\end{eqnarray}
Combining (\ref{aa3}), (\ref{aa4}) and (\ref{aaa}), we have shown that $H({A}_2^{[2]}|W_1, \mathbb{F}, \mathbb{G}) = L/2$. The proof of (\ref{l2}) is complete.

Next we prove (\ref{a12}). On the one hand, 
\begin{eqnarray}
H(A_2^{[2]}|W_1, A_2^{[1]}, \mathbb{F}, \mathbb{G}) \leq H(A_2^{[2]}|W_1, \mathbb{F}, \mathbb{G}) \overset{(\ref{l2})}{=} L/2 \label{a121}
\end{eqnarray}
On the other hand, from sub-modularity of entropy functions we have
\begin{eqnarray}
\lefteqn{H(A_2^{[2]}, A_2^{[1]}|W_1, \mathbb{F}, \mathbb{G}) }\notag\\
&\geq& -H(A_2^{[1]}, A_1^{[1]}|W_1,\mathbb{F}, \mathbb{G}) + H(A_1^{[1]}, A_2^{[2]}, A_2^{[1]}|W_1, \mathbb{F}, \mathbb{G}) + H(A_2^{[1]}|W_1, \mathbb{F}, \mathbb{G}) \\
&\overset{(\ref{eq:by})(\ref{corr})(\ref{a21})}{\geq}& - L/2 + {\color{black} H(A_1^{[1]}, A_2^{[2]}, A_2^{[1]}, W_2|W_1,\mathbb{F}, \mathbb{G}) }  + L/2 \label{sec}\\
&\overset{}{\geq}& H(W_2|W_1, \mathbb{F}, \mathbb{G}) \overset{(\ref{maprandom})}{=} L\\
&\Rightarrow& H(A_2^{[2]}|W_1, A_2^{[1]}, \mathbb{F}, \mathbb{G}) = H(A_2^{[2]},A_2^{[1]}|W_1, \mathbb{F}, \mathbb{G}) - H(A_2^{[1]}|W_1, \mathbb{F}, \mathbb{G}) 
 \overset{(\ref{a21})}{\geq} L/2 \label{a122}
\end{eqnarray}
Note that the second term of (\ref{sec}) follows from the assumption that $A_1^{[1]} = A_1^{[2]}$ so that from $A_1^{[1]}, A_2^{[2]}$, we can decode $W_2$ just as from $A_1^{[2]}, A_2^{[2]}$, we can decode $W_2$.
Combining (\ref{a121}), (\ref{a122}), we have proved $H(A_2^{[2]}|W_1, A_2^{[1]}, \mathbb{F}, \mathbb{G}) = L/2$. Symmetrically, it follows that $H(A_2^{[2]}|W_2, A_2^{[1]}, \mathbb{F}, \mathbb{G}) = L/2$. Therefore, the desired inequality (\ref{a12}) is obtained.
%

\hfill\QED

From Lemma \ref{lemma:id}, we know that $I(A_2^{[1]};A_2^{[2]}|W_1, \mathbb{F}, \mathbb{G}) = I(A_2^{[1]};A_2^{[2]}|W_2, \mathbb{F}, \mathbb{G}) = 0$.  Plugging in Ingleton's inequality \cite{Ingleton} that holds for linear schemes but not for non-linear schemes, we have
\begin{eqnarray}
I(A_2^{[1]};A_2^{[2]}|\mathbb{F}, \mathbb{G}) &\leq& I(A_2^{[1]};A_2^{[2]}|W_1, \mathbb{F}, \mathbb{G}) + I(A_2^{[1]};A_2^{[2]}|W_2, \mathbb{F}, \mathbb{G}) + \underbrace{I(W_1;W_2|\mathbb{F}, \mathbb{G})}_{=0,~\mbox{\footnotesize from}~(\ref{maprandom})} \notag \\
&=& 0 \\
\Rightarrow H(A_2^{[1]}, A_2^{[2]}| \mathbb{F}, \mathbb{G}) &=& H(A_2^{[1]}|\mathbb{F}, \mathbb{G}) + H(A_2^{[2]}| \mathbb{F}, \mathbb{G}) \\
&\overset{(\ref{eq:caps})}{=}& H(A_2^{[1]}|\mathbb{F}, \mathbb{G}) + H(A_1^{[1]}|\mathbb{F}, \mathbb{G}) \\
&\geq& H(A_1^{[1]}, A_2^{[1]}|\mathbb{F}, \mathbb{G}) \\
&\overset{(\ref{corr})}{=}& H(W_1, A_1^{[1]}, A_2^{[1]}|\mathbb{F}, \mathbb{G}) \\
&\overset{}{=}& H(W_1|\mathbb{F}, \mathbb{G}) + H(A_1^{[1]}, A_2^{[1]}|W_1, \mathbb{F}, \mathbb{G}) \\
&\overset{(\ref{maprandom})}{\geq}& L + H(A_1^{[1]}|W_1, \mathbb{F}, \mathbb{G}) \overset{(\ref{l2})}{=} 3L/2
\\
\Rightarrow \alpha = H(S_2)/L &\geq& H(A_2^{[1]}, A_2^{[2]}| \mathbb{F}, \mathbb{G})/L \geq 3/2
\end{eqnarray}

\subsection{Symmetrization}\label{sec:symmetry}
\begin{theorem}\label{thm:symmetry}
Consider the single-round PIR problem with $K = 2$ messages and $N = 2$ databases. Suppose we have a scheme described by $\bar{L},\bar{W}_{1}, \bar{W}_2, \bar{S}_{1}, \bar{S}_2, \bar{Q}_{1:2}^{[1]},  \bar{Q}_{1:2}^{[2]}, \bar{A}_{1:2}^{[1]}, \bar{A}_{1:2}^{[2]}, \bar{\mathbb{F}}, \bar{\mathbb{G}}$. Then we can construct a symmetric PIR scheme, also for $K=N=2$, described by $L,W_{1}, W_2, S_{1}, S_2, Q_{1:2}^{[1]}, Q_{1:2}^{[2]}, A_{1:2}^{[1]}, A_{1:2}^{[2]}, \mathbb{F}, \mathbb{G}$ such that 
\begin{eqnarray}
H(A_1^{[1]}|\mathbb{F}, \mathbb{G}) &=& H(A_2^{[1]}| \mathbb{F}, \mathbb{G}) = H(A_2^{[2]}| \mathbb{F}, \mathbb{G}) \label{e1}\\
H(S_1) &=& H(S_2) \label{e2}\\
L&=&2\bar{L}
\end{eqnarray}
such that the symmetric PIR scheme has the same rate and storage overhead as the original PIR scheme. 
\end{theorem}
{\it Proof:} Consider two independent implementations of the asymmetric PIR scheme. Let us use the `bar' notation for the first implementation and the `tilde' notation for the second implementation. In the first implementation, there are two messages $\bar{W}_1, \bar{W}_2$, each of length $\bar{L}$, two databases $\bar{\mbox{\small DB1}}$ and $\bar{\mbox{\small DB2}}$ which store $\bar{S}_1, \bar{S}_2$, respectively. In the second implementation, there are two messages $\tilde{W}_1, \tilde{W}_2$, each of length $\tilde{L}=\bar{L}$, two databases $\tilde{\mbox{\small DB2}}$ and $\tilde{\mbox{\small DB1}}$ which store $\tilde{S}_1, \tilde{S}_2$, respectively. Note the critical detail that the database indices are switched in the second implementation. The asymmetric PIR scheme specifies the queries for each implementation such that the user can privately retrieve an arbitrarily chosen message from each implementation. 

The symmetric PIR scheme  is created by combining the two implementations.   In the combined scheme, there are two messages $W_1=(\bar{W}_1, \tilde{W}_1)$ and ${W}_2 = (\bar{W}_2, \tilde{W}_2)$, each of length $L=2\bar{L}$, two databases $\mbox{\small DB1}$ and $\mbox{\small DB2}$ which store $(\bar{S}_1,\tilde{S}_2)$ and $(\bar{S}_2,\tilde{S}_1)$, respectively. Retrieval works exactly as before. For example, if the  user wishes to privately retrieve $W_1=(\bar{W}_1, \tilde{W}_1)$, then it retrieves $\bar{W}_1$ exactly as in the first implementation, and $\tilde{W}_1$ exactly as in the second implementation.

Since the symmetric scheme is comprised of two independent implementations of the original PIR scheme, the message size, total download size, total storage size, are all doubled relative to the original PIR scheme. As a result the rate and storage overhead, which are normalized quantities, remain unchanged in the new scheme.  Symmetry is achieved because each database from the original PIR scheme is equally represented within each database in the new PIR scheme.

Mathematically, 
\begin{eqnarray}
{W}_1 = ( \bar{W}_1, \tilde{W}_1), {W}_2 = (\bar{W}_2, \tilde{W}_2) \\
{S}_1 = (\bar{S}_1, \tilde{S}_2), {S}_2 = (\bar{S}_2, \tilde{S}_1) \\
{\mathbb{F}} = (\bar{\mathbb{F}}, \tilde{{\mathbb{F}}}), {\mathbb{G}} = (\bar{\mathbb{G}},\tilde{{\mathbb{G}}}) \\
{Q}_{n}^{[k]} = ( \bar{Q}_{n}^{[k]}, \tilde{Q}_{3-n}^{[k]}), n =1,2, k = 1,2 \\
{A}_{n}^{[k]} = ( \bar{A}_{n}^{[k]}, \tilde{A}_{3-n}^{[k]})
\end{eqnarray}
where each random variable with a bar symbol is independent of and identically distributed with the same random variable with a tilde symbol.
We are now ready to prove the first equality in (\ref{e1}).
\begin{eqnarray}
H(A_1^{[1]}|\mathbb{F}, \mathbb{G}) &=& H(\bar{A}_{1}^{[1]}, \tilde{A}_{2}^{[1]}|\mathbb{F}, \mathbb{G}) \\
&=& H(\bar{A}_{1}^{[1]} |\bar{\mathbb{F}}, \bar{\mathbb{G}}) + H(\tilde{A}_{2}^{[1]} |\tilde{\mathbb{F}}, \tilde{\mathbb{G}}) \label{tosum}\\
&=& H(\tilde{A}_{1}^{[1]} |\tilde{\mathbb{F}}, \tilde{\mathbb{G}}) + H(\bar{A}_{2}^{[1]} |\bar{\mathbb{F}}, \bar{\mathbb{G}}) \label{sameentropy}\\
&=& H(\bar{A}_{2}^{[1]}, \tilde{A}_{1}^{[1]}|\mathbb{F}, \mathbb{G}) \label{tosum2} \\
&=& H(A_2^{[1]}| \mathbb{F}, \mathbb{G}) 
\end{eqnarray}
where $(\ref{tosum})$ and $(\ref{tosum2})$ follow from the fact that the two copies of the given scheme are independent and (\ref{sameentropy}) is due to the property that the two copies are identically distributed. Consider the second equality in (\ref{e1}).
\begin{eqnarray}
H(A_2^{[1]}| \mathbb{F}, \mathbb{G}) &\overset{(\ref{querydet})}{=}& H(A_2^{[1]}| Q_2^{[1]}, \mathbb{F}, \mathbb{G}) \\
&\overset{}{=}& H(A_2^{[1]}| Q_2^{[1]}, \mathbb{G}) \label{mar1} \\
&\overset{(\ref{tprivacy})}{=}& H(A_2^{[2]}| Q_2^{[2]}, \mathbb{G}) \\
&\overset{}{=}& H(A_2^{[2]}| Q_2^{[2]}, \mathbb{F}, \mathbb{G}) \label{mar2}\\
&\overset{(\ref{querydet})}{=}& H(A_2^{[2]}| \mathbb{F}, \mathbb{G})
\end{eqnarray}
where (\ref{mar1}) and (\ref{mar2}) are due to the Markov chain $\mathbb{F} - (Q_2^{[k]},\mathbb{G}) - A_2^{[k]}, k = 1,2$, which is proved as follows.
\begin{eqnarray}
I(A_2^{[k]}; \mathbb{F} | Q_2^{[k]}, \mathbb{G}) &\leq& I(A_2^{[k]}, S_2; \mathbb{F} | Q_2^{[k]}, \mathbb{G})\\
&=& I(S_2; \mathbb{F} | Q_2^{[k]}, \mathbb{G}) + I(A_2^{[k]}; \mathbb{F} | Q_2^{[k]}, \mathbb{G}, S_2)\\
&\overset{(\ref{ansdet})}{=}& I(S_2; \mathbb{F} | Q_2^{[k]}, \mathbb{G}) \\
&\overset{}{\leq}& I(S_2, W_1, W_2; \mathbb{F} | Q_2^{[k]}, \mathbb{G}) \\
&\overset{}{=}& I(W_1, W_2; \mathbb{F} | Q_2^{[k]}, \mathbb{G}) + I(S_2; \mathbb{F} | Q_2^{[k]}, \mathbb{G},W_1, W_2)\\
&\overset{(\ref{storagedet})}{\leq}& I(W_1, W_2; \mathbb{F}, Q_2^{[k]}, \mathbb{G}) \\
&\overset{(\ref{querydet})(\ref{maprandom})}{=}& 0 
\end{eqnarray}
Finally, we prove (\ref{e2}).
\begin{eqnarray}
H(S_1) &=& H(\bar{S}_1, \tilde{S}_2) \\
&=& H(\bar{S}_1) + H(\tilde{S}_2) \label{tosum3}\\
&=& H(\tilde{S}_1) + H(\bar{S}_2) \label{sameentropy2}\\
&=& H(\bar{S}_2, \tilde{S}_1) \label{tosum4}\\
&=& H(S_2) 
\label{same2}
\end{eqnarray}
where $(\ref{tosum3})$ and $(\ref{tosum4})$ follow from the fact that the two copies of the given scheme are independent and (\ref{sameentropy2}) is due to the property that the two copies are identically distributed.
\hfill\QED

\section{Conclusion}\label{sec:conc}
We showed that the capacity of MPIR is equal to the capacity of PIR, both with and without $T$-privacy constraints. Our result implies that there is no advantage in terms of capacity from multiround over single-round schemes, non-linear over linear schemes, or $\epsilon$-error over zero-error schemes. We also offered a counterpoint to this pessimistic  result by exploring optimal storage overhead instead of capacity. Specifically, we constructed a simple multiround, non-linear, $\epsilon$-error PIR scheme that achieves a strictly smaller storage overhead than the best possible with any single-round, linear, zero-error PIR scheme. The simplicity of the scheme makes it an attractive point of reference for future work toward understanding the role of linear versus non-linear schemes, zero-error versus $\epsilon$-error capacity, and single-round versus multiple round communications. Another interesting insight revealed by the scheme is the privacy benefit of reduced storage overhead. By not storing all the information at each database, and by optimally compressing the stored information, not only do we reduce the storage overhead, but also we enable privacy where it wouldn't hold otherwise.

\bibliographystyle{IEEEtran}
\bibliography{Thesis}

\begin{thebibliography}{10}
\providecommand{\url}[1]{#1}
\csname url@samestyle\endcsname
\providecommand{\newblock}{\relax}
\providecommand{\bibinfo}[2]{#2}
\providecommand{\BIBentrySTDinterwordspacing}{\spaceskip=0pt\relax}
\providecommand{\BIBentryALTinterwordstretchfactor}{4}
\providecommand{\BIBentryALTinterwordspacing}{\spaceskip=\fontdimen2\font plus
\BIBentryALTinterwordstretchfactor\fontdimen3\font minus
  \fontdimen4\font\relax}
\providecommand{\BIBforeignlanguage}[2]{{%
\expandafter\ifx\csname l@#1\endcsname\relax
\typeout{** WARNING: IEEEtran.bst: No hyphenation pattern has been}%
\typeout{** loaded for the language `#1'. Using the pattern for}%
\typeout{** the default language instead.}%
\else
\language=\csname l@#1\endcsname
\fi
#2}}
\providecommand{\BIBdecl}{\relax}
\BIBdecl

\bibitem{PIRfirst}
B.~Chor, O.~Goldreich, E.~Kushilevitz, and M.~Sudan, ``Private information
  retrieval,'' in \emph{Proceedings of the 36th Annual Symposium on Foundations
  of Computer Science}, 1995, pp. 41--50.

\bibitem{PIRfirstjournal}
B.~Chor, E.~Kushilevitz, O.~Goldreich, and M.~Sudan, ``{Private Information
  Retrieval},'' \emph{Journal of the ACM (JACM)}, vol.~45, no.~6, pp. 965--981,
  1998.

\bibitem{CPIR}
R.~Ostrovsky and W.~E. Skeith~III, ``{A Survey of Single-database Private
  Information Retrieval: Techniques and Applications},'' in \emph{Public Key
  Cryptography--PKC 2007}.\hskip 1em plus 0.5em minus 0.4em\relax Springer,
  2007, pp. 393--411.

\bibitem{William}
W.~Gasarch, ``{A Survey on Private Information Retrieval},'' in \emph{Bulletin
  of the EATCS}, 2004.

\bibitem{Yekhanin}
S.~Yekhanin, ``{Private Information Retrieval},'' \emph{Communications of the
  ACM}, vol.~53, no.~4, pp. 68--73, 2010.

\bibitem{Sun_Jafar_PIR}
H.~Sun and S.~A. Jafar, ``{The Capacity of Private Information Retrieval},''
  \emph{arXiv preprint arXiv:1602.09134}, 2016.

\bibitem{Sun_Jafar_PIRL}
------, ``{Optimal Download Cost of Private Information Retrieval for Arbitrary
  Message Length},'' \emph{arXiv preprint arXiv:1610.03048}, 2016.

\bibitem{Sun_Jafar_TPIR}
------, ``{The Capacity of Robust Private Information Retrieval with Colluding
  Databases},'' \emph{arXiv preprint arXiv:1605.00635}, 2016.

\bibitem{Sun_Jafar_SPIR}
------, ``{The Capacity of Symmetric Private Information Retrieval},''
  \emph{arXiv preprint arXiv:1606.08828}, 2016.

\bibitem{Banawan_Ulukus}
K.~Banawan and S.~Ulukus, ``{The Capacity of Private Information Retrieval from
  Coded Databases},'' \emph{arXiv preprint arXiv:1609.08138}, 2016.

\bibitem{Wang_Skoglund}
Q.~Wang and M.~Skoglund, ``{Symmetric Private Information Retrieval For MDS
  Coded Distributed Storage},'' \emph{arXiv preprint arXiv:1610.04530}, 2016.

\bibitem{Chan_Ho_Yamamoto}
T.~H. Chan, S.-W. Ho, and H.~Yamamoto, ``{Private Information Retrieval for
  Coded Storage},'' \emph{Proceedings of IEEE International Symposium on
  Information Theory (ISIT)}, pp. 2842--2846, 2015.

\bibitem{dougherty1}
R.~Dougherty, C.~Freiling, and K.~Zeger, ``Insufficiency of linear coding in
  network information flow,'' \emph{IEEE Transactions on Information Theory},
  vol.~51, no.~8, pp. 2745 -- 2759, Aug. 2005.

\bibitem{CG_Duality}
T.~H. Chan and A.~Grant, ``{Dualities between entropy functions and network
  codes},'' \emph{IEEE Trans. Inf. Theory}, vol.~54, no.~10, pp. 4470 -- 4487,
  Oct. 2008.

\bibitem{Rouayheb_Sprintson_Georghiades}
S.~Rouayheb, A.~Sprintson, and C.~Georghiades, ``{On the Index Coding Problem
  and Its Relation to Network Coding and Matroid Theory},'' \emph{IEEE Trans.
  on Inf. Theory}, vol.~56, no.~7, pp. 3187--3195, July 2010.

\bibitem{Blasiak_Kleinberg_Lubetzky_2011}
\BIBentryALTinterwordspacing
A.~Blasiak, R.~Kleinberg, and E.~Lubetzky, ``Lexicographic products and the
  power of non-linear network coding,'' \emph{ArXiv:1108.2489}, Aug. 2011.
  [Online]. Available: \url{http://arxiv.org/abs/1108.2489}
\BIBentrySTDinterwordspacing

\bibitem{Langberg_Effros}
{M. Langberg and M. Effros}, ``{Network Coding: Is zero error always
  possible?}'' in \emph{49th Allerton Conference on Communication, Control and
  Computing.}, 2011, pp. 1478--1485.

\bibitem{Jalali_Effros_Ho}
S.~Jalali, M.~Effros, and T.~Ho, ``On the impact of a single edge on the
  network coding capacity,'' in \emph{Information Theory and Applications
  Workshop (ITA), 2011}.\hskip 1em plus 0.5em minus 0.4em\relax IEEE, 2011, pp.
  1--5.

\bibitem{Kosut_Kliewer}
O.~Kosut and J.~Kliewer, ``On the relationship between edge removal and strong
  converses,'' in \emph{Proceedings of International Symposium on Information
  Theory (ISIT)}, 2016.

\bibitem{Beimel_Ishai}
A.~Beimel and Y.~Ishai, ``{Information-theoretic private information retrieval:
  A unified construction},'' in \emph{Automata, Languages and
  Programming}.\hskip 1em plus 0.5em minus 0.4em\relax Springer, 2001, pp.
  912--926.

\bibitem{Beimel_Ishai_Kushilevitz}
A.~Beimel, Y.~Ishai, and E.~Kushilevitz, ``{General constructions for
  information-theoretic private information retrieval},'' \emph{Journal of
  Computer and System Sciences}, vol.~71, no.~2, pp. 213--247, 2005.

\bibitem{Shah_Rashmi_Kannan}
N.~Shah, K.~Rashmi, and K.~Ramchandran, ``{One Extra Bit of Download Ensures
  Perfectly Private Information Retrieval},'' in \emph{Proceedings of IEEE
  International Symposium on Information Theory (ISIT)}, 2014, pp. 856--860.

\bibitem{Fazeli_Vardy_Yaakobi}
A.~Fazeli, A.~Vardy, and E.~Yaakobi, ``{Codes for distributed {PIR} with low
  storage overhead},'' in \emph{Proceedings of IEEE International Symposium on
  Information Theory (ISIT)}, 2015, pp. 2852--2856.

\bibitem{Tajeddine_Rouayheb}
R.~Tajeddine and S.~E. Rouayheb, ``{Private Information Retrieval from MDS
  Coded Data in Distributed Storage Systems},'' \emph{arXiv preprint
  arXiv:1602.01458}, 2016.

\bibitem{Rao_Vardy}
S.~Rao and A.~Vardy, ``{Lower Bound on the Redundancy of PIR Codes},''
  \emph{arXiv preprint arXiv:1605.01869}, 2016.

\bibitem{Blackburn_Etzion}
S.~Blackburn and T.~Etzion, ``{PIR Array Codes with Optimal PIR Rate},''
  \emph{arXiv preprint arXiv:1607.00235}, 2016.

\bibitem{Blackburn_Etzion_Paterson}
T.~E. Simon R.~Blackburn and M.~B. Paterson, ``{PIR schemes with small download
  complexity and low storage requirements},'' \emph{arXiv preprint
  arXiv:1609.07027}, 2016.

\bibitem{Zhang_Wang_Wei_Ge}
Y.~Zhang, X.~Wang, H.~Wei, and G.~Ge, ``{On private information retrieval array
  codes},'' \emph{arXiv preprint arXiv:1609.09167}, 2016.

\bibitem{Slepian_Wolf}
D.~Slepian and J.~Wolf, ``Noiseless coding of correlated information sources,''
  \emph{IEEE Transactions on information Theory}, vol.~19, no.~4, pp. 471--480,
  1973.

\bibitem{Cover_Thomas}
T.~M. Cover and J.~A. Thomas, \emph{Elements of Information Theory}.\hskip 1em
  plus 0.5em minus 0.4em\relax Wiley, 2006.

\bibitem{Ingleton}
A.~W. Ingleton, ``{Representation of matroids in combinatorial mathematics and
  its applications},'' \emph{Combinatorial Mathematics and Its Applications},
  vol.~44, pp. 149 -- 167, Jul. 1971.

\end{thebibliography}
\end{document}